Title: Operads and Quantum Gravity
Authors: I.P. Zois
Comments: Replacement of hep-th/0302114, new extended version accepted for publication, added references
Report-no:
Journal-ref:To appear in Repts Math. Phys. 55.3 (2005). 
Initial version appeared as poster/oral presentation at XIVth ICMP 2003 Lisbon.

We discuss the use of the language of operads to try to understand holography and through that quantum gravity. 
As we find out, the Deligne Conjecture and the action of the Grothendieck-Teichmuller group play an important role.

\documentstyle[11pt,amsfonts,amssymb,amscd]{article}
\begin{document}
\title{Operads and Quantum Gravity}
\author{Ioannis P. \ ZOIS\thanks{izois@ihes.fr Current address: 
School of Mathematics, Cardiff University, PO Box 926, Cardiff CF24 4YH, UK; 
e-mail address: zoisip@cf.ac.uk}\\
\\
IHES, Le Bois-Marie, 35, route de Chartres
\\
F-91440 Bures-sur-Yvette, FRANCE
\\
and
\\
Mathematical Institute, Oxford University
\\
24-29 St. Giles', Oxford, OX1 3LB, UK}
\date{}

\maketitle

\begin{center}
\emph{``In girum imus nocte et consumimur igni''}.\\
\end{center}
\begin{abstract}

In this article we try to explain and extend a statement due
to Maxim Kontsevich back in 1999, that the \emph{Holography Principle}
in physics should be related to the (higher dimensional)
\emph{Deligne Conjecture} in mathematics. This seems to suggest that the
\emph{little $d$-discs operad} (or equivalently the notion of a
\emph{$d$-algebra}) gives a new way to understand the mathematical aspects of
\textsl{quantum gravity} using \textsl{holography}. The strategy is as
follows: we would like
to learn something about quantum gravity in $(d+1)$ dimensions: we use
holography to reduce our original problem to a CFT in $d$-dimensions. The
deep origin of this dimensional reduction lies on the fact that it is the
\textsl{area} and \emph{not} the \textsl{volume} which appears in the
formula giving the entropy of
black holes as described long ago by Hawking. Then we use $d$-algebras (i.e. 
the little $d$-discs operad) to study our $d$-dim CFT.
The possible relation between $d$-dim CFT and $d$-algebras comes from the 
lesson we have learnt from strings (namely the 2-dim CFT case): the space of 
physical states
in closed string field theory (ie the BRST cohomology) has a natural
\emph{Gerstenhaber} algebra structure and this by Cohen's theorem is related to
the little 2-discs operad. The proposal then is that the relation might hold 
in higher
than 2 dimensions. This approach is algebraic although it would have been 
much more satisfactory if we could generalise Segal's geometric approach
to CFT in higher than 2 dimensions. Hopefully the article is mathematically
self-contained.


\end{abstract}

\section{Introduction}

This work was motivated by an attempt to understand an interesting
statement
by Maxim Kontsevich \cite{kon1} back in 1999: that the \emph{Holography
Principle} originally due to G. 't Hooft \cite{thooft} in
Physics might be related to the \emph{Higher Dimensional Deligne
Conjecture} in Mathematics due to Kontsevich. In 2000 Kontsevich and
Soibelman \cite{kon2} proved the original Deligne conjecture. A secong proof was given by Tamarkin (see \cite{tamarkin}).
Very recently yet another proof of the Deligne conjecture appeared which is due to A. Kohlhuber 
using the \emph{arc operad}.\\

Let us roughly explain both these statements here and elaborate more on them
later: Holography is a statement about \emph{quantum gravity}; in
simple terms it says that quantum gravity must be a \textsl{topological
quantum field theory}.\\

On the other hand the original Deligne conjecture (due to Deligne as the name 
suggests, see \cite{deligne}), it is about the
Hochschild complex of associative algebras; as it is well-known the
Hochschild complex is very useful when one wants to study the theory of
deformations of algebras.\\

We organise this article as follows: in section 2 we try to explain the
necessity for holography in physics starting from black hole puzzles; in
section 3 we give the necessary
mathematical definitions about operads and Gerstenhaber algebras
and we also state the two basic theorems: the first is due to F. Cohen and
relates Gerstenhaber algebras with the little 2-discs operad and the second
is D. Tamarkin's theorem on the formality of the little $d$-discs operad.
In the last section we explain the relation between string theory and the
little 2-discs operad.
Then as we shall see, one will be able to interpret Kontsevich's statement as
a higher dimensional analogue of the above relation.\\

We start our discussion with possibly the most mysterious objects
in the universe: \emph{Black Holes.}

\section{Understanding Black Holes from Strings: the need for a nonlocality
mechanism}

We have gained some understanding on two
important problems in
black hole physics (abreviated to ``BH'' in the sequel) by using
some recent results from string theory dualities (for more details one
can see \cite{das} which is a nice review article):\\

{\bf 1.}  In general relativity we have the
so-called \emph{``no hair''} theorem
which refers to black holes. This is the statement that the configuration
of a black hole solution given by the Schwarzchild metric
is \emph{uniquely determined} by its mass (= total energy) (we assume
no more conserved quantities like electric charge or angular momentum for
simplicity). In other words we have only {\bf one} configuration (for a given
mass) associated to the Schwarzchild solution.

Following the usual definition for the classical \emph{entropy} of a system
$$S=k_{B}ln\Omega $$ where $\Omega $ is the number of microstates compatible
with some given values of the macroscopic parameters (eg temperature,
pressure, volume etc) and $k_{B}$ is Boltzmann's constant, we
immediately deduce that a BH must have \emph{zero} entropy classically since
$ln1=0$.\\

But then we encounter the \textsl{qualitative argument} originally due
to Beckenstein that if this was indeed the case, then any object, e.g. some
gas, falling into a BH would \textsl{contradict the second law of
thermodynamics}. To avoid that one should associate a \emph{nonzero} entropy
(positive of course) to any BH. The precise value of the
entropy $S$ of a BH was then determined by the Hawking area formula which,
ignoring constants, reads (see for example \cite{cam})
$$S\sim A$$
where $A$ is the \emph{area} of the event horizon. So if we denote by $R$ the radius of the event horizon of a BH,
its entropy is proportional to $R^{2}$ and \textsl{not proportional to} $R^{3}$. The origin of BH entropy
was understood to be \emph{quantum mechanical}.\\

At this point we should make a comment: the Hawking formula is quite 
surprising, remarkable and counterintuitive since from what we know from 
thermodynamics the entropy of a system depends on its \emph{volume} but in this
 case of black holes a \textsl{dimensional reduction} occures and in fact the 
entropy is proportional only to the \emph{area} of its event horizon.\\

Nonetheless we know that statistical physics gives a more fundamental 
explanation of the laws of thermodynamics and moreover a correct ``would-be'' 
quantum theory of gravity should explain the origin of the quantum states 
associated to a BH. So the challenge was to find a statistical explanation for 
the quantum states associated to a BH which give rise to its entropy described 
by the Hawking formula.\\

\emph{Superstring theory} can indeed, in some cases, provide an explanation 
for the origin of
quantum states associated to \emph{multicharged extremal black holes}. And
anyway string theory is arguably the best known
candidate for a quantum theory of gravity. The argument
 which explains the microscopic origin of BH entropy starting from string
theory was originally due to Strominger, Vafa, Horowitz and Maldacena and it 
is based
on {\bf S}-duality. The later is a statement about an isomorphism between
\emph{strong} and \emph{weak} coupling regions of superstring theory;
equivalently it \textsl{interchanges} \emph{monopoles} with \emph{charges} in
the theory (or equivalently it interchanges topology and dynamics) and gives
us the ability to identify \textsl{BPS superstring states} which will either
be
\textsl{perturbative states} if they carry \emph{NS charges} or
\textsl{D-branes} if they carry \emph{R charges} in
\emph{``weak coupling''} region with \emph{extremal black holes} carrying the
analogous type of charge in \emph{``strong coupling''} region. We restrict
our attention to BPS states (these are states whose mass does not receive any
quantum corrections) because for simplicity we assume no backcreation for the
black hole (namely its mass which is equal to its energy remains constant).
Briefly then the main
idea behind this string theoretic explanation of the quantum states
associated to a black hole is that since superstrings live in 10-dim and BH in
4-dim, the remaining
6 compactified dimensions essentially provide a ``phase space'' which we
quantize and thus we obtain the states of the BH. (This picture is not 
utterly correct but we think captures the spirit of the argument
and gives a clear picture conceptually). For a possible noncommutative generalisation of the Beckenstein-Hawking 
area-entropy formula for BHs see \cite{z2}.\\

{\bf 2.}  The second problem we would like to consider in BH physics is the
so-called \emph{``information paradox''}. \emph{Classically},
\textsl{nothing can escape the event horizon of a BH} (since that
would require a velocity which is grater than the velocity of light; one can
use that as a definition of the \emph{``event horizon''}). Yet
quantum mechanically, since a BH has a positive entropy as we just
argued above, \emph{assuming a thermodynamical behaviour}, it should also
have a corresponding \emph{temperature} from the well-known relation in
thermodynamics
$$(dM=)dE=TdS.$$
This is the \emph{Hawking temperature}
$$T_{H}=\frac{\hbar \kappa }{2k_{B}\pi }$$
where $\kappa $ is \textsl{surface gravity} (the acceleration felt by a
static object at the horizon as measured from the asymptotic region),
and hence BH's should also \emph{radiate.}\\

This is the \emph{Hawking radiation}. Then the problem with
radiation carrying out the information of \emph{formation}, assuming
\textsl{no quantum xeroxing,} is to maintain \textsl{unitarity} of the process
as ordinary quantum mechanics requires. More precisely, the way one computes
Hawking radiation is by assuming a \emph{codim-1 foliation} of spacetime
where the normal direction is time
and this radiation process appears to be \emph{nonunitary}. (For a more general discussion about foliations 
and the relation between physical and topological entropy one can see \cite{z1} and \cite{z2}).\\

Thus trying to avoid a classical contradiction with the 2nd law of
thermodynamics we assumed that BH's have positive entropy (whose origin is
quantum mechanical); yet this almost immediately created another contradiction
with quantum mechanics: loss of unitarity in BH radiation. Unitarity is
absolutely crucial in any quantum theory since it reflects the
conservation of probabilities. So it seems that we didn't actually achieve
very much: we simply \emph{``pushed''} the contradiction from the classical
to the quantum realm.\\

It appears that the most \emph{economical} (i.e. requiring the fewest
changes to things we already know in physics) \textsl{way out} is to assume
that there is some \textsl{physical principle} which does not
allow this to happen. We \emph{enforce} \textsl{unitarity} throughout by
imposing a \textsl{non-locality} mechanism. One such mechanism is the
\emph{holography principle} due to G. 't Hooft (see \cite{thooft}). The original statement is the
following:\\

\emph{``A quantum theory of gravity on a
$(d+1)$-manifold with boundary should be equivalent to a conformal field
theory (CFT for short) on the boundary (which is a $d$-manifold) and this
conformal field
theory on the boundary must have one degree of freedom per Planck area''}.\\

Let us elaborate more on this: since for BH we seem to lose any information
passing the event horizon, it is reasonable to assume that in order to avoid
this problem (along with its quantum mechanical incarnation of nonunitary
radiation), everything that happens
inside the black hole should be described from data on its event horizon. It
 is clear we think that the motivation for this \textsl{dimensional reduction}
of quantum gravity in holography came from the formula for the black hole
entropy: the entropy of a black hole is proportional to the \emph{area} and 
\textsl{not} the volume of the event horizon.\\

Another very useful way of thinking about the holography principle is that
it simply says that for a given 3-volume $V$
in space the \emph{state of maximal entropy} in nature is given by the
\textsl{largest BH that fits inside $V$},
(silently we are making use of the Hawking formula which says that the
entropy of a BH is
proportional to the area of its event horizon).\\

This principle has a deep consequence on \emph{perturbative quantum field
theory:} BH's
provide a \textsl{natural} \emph{cut-off limit} since the
above statement says that for fermions for example one cannot have a huge
amount of energy concentrated
in a tiny region of space because that would collapse into a BH.\\

There is also a superstring theoretic version
of the holography principle using string theory language,
the so-called \emph{``Maldacena
conjecture''} which states that string theory on the smooth manifold
$AdS_{5}\times S^{5}$ is dual to $N=4$ SYM $SU(N)$ gauge theory on the
\emph{boundary} of $AdS_{5}$. In this talk we shall primarily
build our understanding
of Kontsevich's statement based on this string theoretic version of
holography. However let us for the moment go back to the original 't Hooft
version of holography and ask:\\

{\bf (Key Question:)} What is a $D=d$ CFT?\\

In order to answer the above question and eventually understand Kontsevich's
statement, we should
start by trying to understand the $D=2$ case first. We know from G. Segal
what a $D=2$ CFT is, so it seems that somehow we have to generalise his work.\\

A good motivation to study $D=2$ CFT comes from string theory itself, one can
say in fact that $D=2$ CFT is intimately related to string theory: strings
are 1-dim objects which in time sweep out 2-manifolds called
\emph{worldsheets};
this is the higher dimensional analogue of the paths (1-dim geometric objects)
swept out in time by point particles. However now we are talking about
M-Theory in physics which generalises string theory and M-Theory contains
the M2 and M5 branes; these are 2 and
5-dim objects respectively whose worldsheets are 3 and 6-dim manifolds. So
apart from holography, there is additional motivation coming from M-Theory to
understand higher dimensional CFT's: a $D=d$ CFT is the theory describing $(d-1)$-branes.\\

Now we would like to describe briefly what string theory is \emph{classically}:
basically it is a $\sigma $\textsl{-model}, namely it describes
\emph{harmonic maps} $\phi :\Sigma _{g}\rightarrow X^{10}$
where $\Sigma _{g}$ is a Riemann surface of genus $g$ representing the
worldsheets of strings and $X$ is a 26-dim
Riemannian manifold with a $B$-field. The dimensionality of $X$ is fixed from
consistency arguments (compatibility with special relativity and cancellation
of the conformal anomaly). The $B$-field is a real valued 2-form
which is used as a potential to \emph{gauge} the worldsheets of the strings
in order to get our Dirac phase factors; in fact one can think of it as the
Poincare Dual (a 2-form) of the worldsheet which is a 2-manifold (it is the
analogue of the gauge
potentials in Yang-Mills theory, connection 1-forms, although now it has to
be a 2-form instead of a 1-form since we are talking about strings whose
worlsheets are 2-manifolds whereas for point particles we needed 1-forms
because their ``worlsheets'' were 1-dim objects). This picture needs to include
fermions as well in order to be complete but we shall not elaborate more on
this. The introduction of fermions along with supersymmetry reduces the
dimensionality of $X$ from 26 down to 10.\\

The \emph{quantum} theory of strings is essentially a \emph{$D=2$ CFT}
(plus a little bit
more structure as we shall see later). In order to
describe $D=2$ CFT one may use the original \emph{geometric} approach due to
G. Segal. This approach however does not lead to the statement of
Kontsevich in a straightforward way. Moreover it is not easy to see how it
can be generalised to
higher dimensions which is what we are after. Instead we shall adopt
 an \emph{algebraic approach} using the language of \emph{operads};
in fact we shall see the \emph{little 2-discs operad} $C_{2}(n)$ arising
naturally in our discussion. This is the crucial step in
order to understand Kontsevich's statement which is the higher
dimensional version of this beautiful fact. The appearence of the operad
$C_{2}(n)$ in string theory is not at all obvious and at least for us quite
surprising. The link between string theory and the little 2-discs operad
$C_{2}(n)$ comes from a deep theorem due to Fred Cohen as we shall try to
exhibit shortly. For simplicity we shall restrict our discussion to
\emph{closed strings}. But we shall do that in the last section because we
need some mathematical definitions first. The final remark here is that we
would still like to generalise Segal's work and get a geometric definition of
$D=d$ CFT. Currently this seems out of reach since there are two reasons
which make Segal's approach particularly nice for the $D=2$ case (but at the
same time act as
barriers when trying to generalise into higher dimensions): the
conformal group in this case is infinite dimensional and hence contains a lot
more information whereas in higher dimensions the conformal group is only
finite dimensional. The second reason is that
the classification of (compact say) 2-manifolds is simple: compact 2-manifolds 
are classified by their genus (ie the number of holes) whereas for $D=3$ it is 
not known if 3-manifolds can be classified and in dimensions $D\geq 4$ we can
only classify simply connected manifolds (in perturbative quantum field theory
that means we can only talk about tree level).\\

\section{A Mathematical Interlude}

In this section now we shall give formal definitions.\\

{\bf Definition 1:}\\
A \emph{Gerstenhaber algebra} (or a $G$-algebra) is a
graded vector space $V=\oplus _{i\in {\bf Z}}V_{i}$ with a \emph{dot product}
$x\cdot y$ defining the structure of a \textsl{graded commutative associative
algebra} along with a \emph{bracket operation} $[x,y]$ of degree -1 defining
the structure of a \textsl{graded Lie algebra} such that the bracket is a
\emph{derivation} with respect to the dot product, i.e. it satisfies the
Leibniz rule
$$[x,y\cdot z]=[x,y]\cdot z+(-1)^{(deg(x)-1)deg(y)}y\cdot [x,z]$$

{\bf Examples:}\\
{\bf i.} Let $A$ be an associative algebra and let
$C^{*}(A,A)$ be its Hochschild complex
where $C^{i}(A,A):=Hom(A^{\otimes i},A)$ and the differential $d$ is defined
as ($x\in C^{n}$):
$$(dx)(a_{1}\otimes ...\otimes a_{n+1}):=a_{1}x(a_{2}\otimes ...\otimes a_{n+1})+\sum _{i=1}^{n}x(a_{1}\otimes ...\otimes a_{i}a_{i+1}\otimes ...\otimes a_{n+1})+...$$
$$+(-1)^{n+1}x(a_{1}\otimes ...\otimes a_{n})a_{n+1}$$
On the Hochschild complex we can define the usual cup product
$$\cup :C^{k}\otimes C^{l}\rightarrow C^{k+l}$$ as follows
($x\in C^{k}$, $y\in C^{l}$ and $a_{i}\in A$):
$$(x\cup y)(a_{1}\otimes ...\otimes a_{k+l}):=(-1)^{kl}x(a_{1}\otimes ...\otimes a_{k})y(a_{k+1}\otimes ...\otimes a_{k+l})$$
Moreover we can also define the \emph{Gerstenhaber bracket} $[,]:C^{k}\otimes C^{l}\rightarrow C^{k+l-1}$ as:
$$[x,y]:=x\circ y-(-1)^{(k-1)(l-1)}y\circ x$$
where
$$(x\circ y)(a_{1}\otimes ...\otimes a_{k+l-1}):=\sum _{i=1}^{k-1}(-1)^{i(l-1)}x(a_{1}\otimes ...\otimes a_{i}\otimes y(a_{i+1}\otimes ...\otimes a_{i+l})\otimes ...\otimes a_{k+l-1})$$
The G-bracket gives after a shift of the ${\bf Z}$-grading by -1 the structure 
of a (differentiable graded lie algebra) DGLA on the Hochschild complex. The 
cup product is not graded commutative (it is only associative) but the induced 
operation on cohomology is graded commutative. Moreover the G-bracket induces 
an operation on cohomology which satisfies the Leibniz rule with respect to 
the cup product, hence the \emph{Hochschild cohomology of any associative 
algebra is in fact a $G$-algebra}.\\

We shall briefly mention three more examples of $G$-algebras:\\

{\bf ii.} \emph{Polyvector fields} on smooth manifolds with \textsl{wedge
product} and \textsl{Schouten-Nijenhuis bracket}.\\

{\bf iii.} \emph{Exterior algebra of a Lie algebra} with \textsl{wedge
product and extension of the Lie bracket.}\\

{\bf iv.} \emph{(Rational) homology of double loop space} with
\textsl{Pontrjagin product and Samelson bracket}.\\

{\bf Definition 2:}\\
An \emph{operad} $P$ (of vector spaces) consists of the following data:\\
{\bf a.} a collection of vector spaces $P(n), n\geq 0$,\\
{\bf b.} an action of the symmetric group $S_{n}$ on $P(n)$ for every $n$,\\
{\bf c.} an identity element $id_{P} \in P(1)$,\\
{\bf d.} compositions $m_{(n_{1},...,n_{k})}$
$$P(k)\otimes (P(n_{1})\otimes  ... \otimes P(n_{k}))\rightarrow P(n_{1}+...+n_{k})$$
for every $k\geq 0$ and $n_{1},...,n_{k}\geq 0$. These compositions have to be
 \emph{associative}, \emph{equivariant} with respect to the symmetric group 
actions and the \emph{identity} element $id_{P}$ has to satisfy the following 
naturality property with respect to the composition:
$$m_{(n)}(id_{P},p)=p$$
and
$$m_{(n,1,...,1)}(p,id_{P},...,id_{P})=p$$
for all $p\in P(n)$ (one can have a look at \cite{operads} for more details).\\

{\bf Example:} The \emph{``endomorphism operad''} of a vector space $V$ is
given
by $P(n):=Hom(V^{\otimes n}, V)$ where the action of the symmetric group and
the identity element are the obvious ones and the compositions are defined by
the substitutions
$$(m_{(n_{1},...,n_{k})}(\phi \otimes (\psi _{1}\otimes ... \otimes \psi _{k}))) (v_{1} \otimes ... \otimes v_{n_{1}+...+n_{k}})$$
$$:=\phi (\psi _{1}(v_{1}\otimes ... \otimes v_{n_{1}})\otimes ... \otimes \psi _{k}(v_{n_{1}+...+n_{k-1}+1}\otimes ...$$
$$ \otimes v_{n_{1}+...+n_{k}}))$$
where $\phi \in P(k):=Hom(V^{\otimes k}, V)$, $\psi _{i}\in P(n_{i}):=Hom(V^{\otimes n_{i}}, V)$ and $i=1,2,...,k$.\\

{\bf Definition 3:}

An \emph{algebra} \textsl{over an operad} $P$ (of vector spaces), or a
\emph{P-algebra}, (or equivalently a \emph{representation} of the operad $P$), 
consists of a vector space $A$
and a collection of multilinear maps $f_{n}:P(n)\otimes A^{\otimes n}
\rightarrow A$ for all $n\geq 0$ satisfying the following axioms:\\
{\bf a.} for any $n\geq 0$ the map $f_{n}$ is $S_{n}$-equivariant,\\
{\bf b.} for any $a\in A$ we have $f_{1}(id_{P}\otimes a)=a$,\\
{\bf c.} all compositions in $P$ map to compositions of multilinear
operations in $A$.\\

In other words the structure of an algebra over $P$ on a vector space $A$ is
given by a \emph{homomorphism} from $P$ to the \emph{endomorphism operad of
$A$}.\\
One can also define \emph{modules} over algebras over operads.\\

One can construct operads denoted $Assoc(n)$, $Lie(n)$, $Poisson(n)$, 
$G(n)$,
$A_{\infty }(n)$,
$L_{\infty }(n)$, $G_{\infty }(n)$
such that algebras over these operads are \emph{associative} algebras,
\emph{Lie} algebras, \emph{Poisson} algebras, \emph{Gerstenhaber} algebras 
and their \emph{homotopic} versions
respectively but there is {\bf no} operad for \emph{Hopf algebras} (for more 
details see \cite{operads}).\\

In the definition of operads we can replace our
\textsl{vector space} $V$ by a compact \emph{topological space} $X$ and hence
define \emph{operads over topological spaces} replacing
\textsl{tensor product} with \textsl{Cartesian product}.\\

The analogue of the endomorphism operad will in this case be
$P(n):=\{Continuous Maps:X^{n}\rightarrow X\}$. Then one can define algebras
over topological operads accordingly.\\

More generally one can define operads and algebras over operads over objects
of any \emph{symmetric monoidal category} $\mathcal C$, namely a category
endowed with the functor $\otimes :\mathcal C\times \mathcal C\rightarrow
\mathcal C$, the identity element $1_{\mathcal C}\in Obj(\mathcal C)$ and the
appropriate coherence isomorphisms for associativity and commutativity of
$\otimes $-product.\\

Operads themselves can be seen as algebras over the \emph{coloured operad.}
(see \cite{kon2} p.12 Remark 1.)\\

In particular we would like to consider operads in the symmetric monoidal
category $Complexes$ whose objects are {\bf Z}-graded complexes of abelian
groups and arrows morphisms of complexes. These are called \emph{differential
graded operads} or \emph{dg-operads} for short. So each component $P(n)$ of an
operad of complexes will be a complex, namely a vector space decomposed into
a direct sum $P(n)=\oplus _{i\in {\bf Z}}P(n)^{i}$ and endowed with a
differential $d:P(n)^{i}\rightarrow P(n)^{i+1}$ of degree $+1$ such that
$d^{2}=0$. Then every dg-operad $P$ has a corresponding \emph{homology operad}
denoted $H_{*}(P)$.\\

{\bf Key idea:}

There is a natural way to construct an \emph{operad of complexes} from
a \textsl{topological operad} by using essentially the
\emph{singular chain complex} of topological spaces.

Let $d\geq 1$ be an integer. Denote by $G_{d}$ the $(d+1)$-dimensional Lie
group of \emph{affine transformations} acting on ${\bf R^{d}}$ via
 $u\mapsto \lambda u+v$ where $\lambda >0$ is a real number and
$v\in {\bf R^{d}}$ is a vector.
This group acts simply transitively on the space of closed discs in
${\bf R^{d}}$ and the disc with centre $v$ and radius $\lambda $ is obtained
from the standard disc
$$D_{0}:=\{(x_{1},...,x_{d})\in {\bf R^{d}}|x_{1}^{2}+...+x_{d}^{2}\leq 1\}$$
by a transformation from $G_{d}$ with parameters $(v,\lambda )$.\\

{\bf Definition 4:}

The \emph{little d-discs operad} $C_{d}(n)$ is a topological operad with the
following structure:\\
{\bf a.} $C_{d}(0):=\emptyset $,\\
{\bf b.} $C_{d}(1):=*$,\\
{\bf c.} for $n\geq 2$ the space $C_{d}(n)$ is the \emph{space of
configurations
of $n$ disjoint d-discs} $(D_{i})_{1\leq i\leq n}$ inside the standard $d$-disc
$D_{0}$.\\

The composition
$$C_{d}(k)\times C_{d}(n_{1})\times ...\times C_{d}(n_{k})\rightarrow C_{d}(n_{1}+...+n_{k})$$
is obtained by applying elements from $G_{d}$ associated with discs
$(D_{i})_{1\leq i\leq n}$ in the configuration in $C_{d}(k)$ to configurations
in all $C_{d}(n_{i})$, $i=1,2,...,k$ and putting the resulting configurations
together. The action of the symmetric group $S_{n}$ on $C_{d}(n)$ is given
by renumerations of indices of discs  $(D_{i})_{1\leq i\leq n}$.\\

{\bf Remark:} The \emph{little $d$-discs operad} $C_{d}(n)$ was introduced by
Peter May (see \cite{may}) and Boardmann-Vogt (see \cite{vogt}), in the late 70's in order
to describe \emph{homotopy types} of \emph{$d$-fold loop spaces}, namely spaces
of continuous maps
$$Maps(S^{d}_{+}, X_{+})$$
where ``$+$'' denotes base point, $X$ is a topological space, $S^{d}$ is the 
$d$-dim sphere. The little $d$-discs operad is the most important operad in
homotopy theory.\\

The key result relating the little $d$-discs operad $C_{d}(n)$ with $d$-fold
loop spaces is that (with field coefficients) chains of $d$-fold loop spaces
become naturally $d$-algebras i.e. algebras over the operad
$Chains C_{d}(n)$. (In fact the above statement is true even without taking
``chains'' in both sides).\\

We have the following well-known\\
{\bf Fact:} The space $C_{d}(n)$ is \textsl{homotopy equivalent} to the
\emph{configuration space of $n$ pairwise distinct points} in ${\bf R^{d}}$:
$${\bf F}(n,{\bf R^{d}}):=({\bf R^{d}})^{n}-Diag=\{(v_{1},...,v_{n})\in ({\bf R^{d}})^{n}$$
$|v_{i}\neq v_{j}$ for $i\neq j\}$\\

{\bf Definition 5:}

Let $\tilde{C_{d}(n)}:={\bf F}(n,{\bf R^{d}})/G_{d}$ which is also the
\emph{Fulton-MacPherson operad} $FM_{d}(n)$.\\

For $n=2$, $FM_{d}(2)$ is homotopy equivalent to the $(d-1)$-sphere $S^{d-1}$.
For all $n\geq 3$, $FM_{d}(n)$ is a \emph{manifold with corners} which can be
identified explicitly.\\

{\bf Definition 6:}

For $d\geq 0$, a $d$-algebra is an algebra over the operad
$Chains (C_{d})$ in the category of complexes.\\

One then has:\\

{\bf Theorem 1.} (F. Cohen, see \cite{cohen}).

There is a natural homotopy equivalence
$$G(n)\simeq H_{*}[Chains C_{2}(n)]$$

Recall the fact that the Hochschild cohomology of any associative algebra has
a natural $G$-algebra structure. The original {\bf Deligne conjecture} (see \cite{deligne}), was
that the
Hochschild complex of an associative algebra (or more generally the Hochschild
complex of an $A_{\infty }$-algebra as mentioned in \cite{kon2}) itself
carries a natural 2-algebra structure, i.e. it has an action of the operad
$Chains C_{2}(n)$.
Its higher dimensional version due to Kontsevich says:\\

\emph{``For any $d-algebra$ there is a natural action of a
universal (in an appropriate sense defined up to homotopy) $(d+1)-algebra$''}.\\

Useful facts:
$$Assoc(n)\simeq H_{*}[Chains C_{1}(n)]$$
$$Lie(n)\simeq H_{n-1}[Chains C_{2}(n)]$$
Since in general homotopic versions of various algebras appear when the 
product is originally defined on the cohomology and one wants to ``lift'' the 
structure to the cochain level,  one has the following general relations 
between algebras and their
``homotopic versions'':
$$Assoc(n)\simeq H_{*}[A_{\infty }(n)]$$
$$Lie(n)\simeq H_{*}[L_{\infty }(n)]$$
$$G(n)\simeq H_{*}[G_{\infty }(n)]$$

{\bf Aside:} The above discussion was about the little discs operad and 
based loop spaces.
One also has a variation of the above, the so called \emph{framed} little 
$d$-discs operad denoted $C^{f}_{d}(n)$
which is related to \emph{free} loop spaces. The framed little 2-discs operad
$C_{2}^{f}(n)$ is \emph{homotopic} to the \textsl{cactus} operad
and the (rational) homology of the framed little 2-discs operad is homotopic 
to the BV-operad. The main result is then that the (rational) homology of the 
free loop 
space $H_{*}(LX)$ where $X$ is a compact oriented manifold (after an
apropriate shift) has a BV-algebra structure, namely it is an algebra over 
the \emph{Batalin-Vilkovisky} operad. At this point we would like to remind 
the reader of the fact that \emph{BV-algebras} appear in the 
\textsl{Lagrangian formalism} of 
field theories whereas \emph{Gerstenhaber algebras} appear in 
BRST cohomology which is 
\textsl{Hamiltoniam formalism} of a field theory. For more details on the 
cactus operad, free loops and BV algebras we refer to \cite{oper}.\\

Then the second important result is the following\\

{\bf Theorem 2:} (D. Tamarkin 1998, see \cite{tamarkin}).

In characteristic zero, the operad $Chains (C_{d})\otimes {\bf R}$ is
{\bf formal}, i.e. it
is homotopy equivalent to its corresponding homology operad.

\section{The appearence of the operad $C_{2}(n)$ in string theory}

After giving all these mathematical definitions we now return back to physics.
G. Segal (see \cite{segal}), defined a $D=2$ CFT as roughly a topological vector space $H_{S^{1}}$
and to each \emph{cobordism}
(which physically represents a Feynman
diagram for strings which are 1-dim objects with 2-dim worldsheets)
$\Sigma _{g}:S^{1}\rightsquigarrow S^{1}$ we associate an operator
$U_{\Sigma _{g}}$ where $\Sigma _{g}$ is a Riemann surface of genus $g$ with a 
conformal class on it. We can replace the boundary of $\Sigma _{g}$ with more 
copies of $S^{1}$ representing more than one incoming and outgoing closed 
strings. The case $g=0$ corresponds to \emph{tree-level} in physics.\\

The space of maps $U_{\Sigma _{g}}$ is parametrised by 
$A:=Conf(\Sigma _{g})/Diff(\Sigma _{g},\partial \Sigma _{g})$ and $A$ acts on
$H_{S^{1}}$ where the denominator denotes diffeomorphisms which become the 
identity on the boundary. $A$ is only a semigroup under concatenation but
it has a Lie algebra $Vect(S^{1})_{\bf C}$ and its fundamental group is 
${\bf Z}$. We have a composition law on $H_{S^{1}}$ for every conformal 
structure on $\Sigma _{g}$ which is associative up to the action of $A$.\\

More concretely we use Lagrangian formalism and write

$$U_{\Sigma _{g}}=\int _{\phi :\Sigma _{g}\rightarrow {\bf R}}e^{-S(\phi )}\mathcal{D}\phi $$

where

$$S(\phi ):=\frac{1}{2}\int _{\Sigma _{g}}d\phi \wedge *d\phi $$

If we change the conformal class on $\Sigma _{g}$ by $f$, then 
$U_{\Sigma _{g}}$ is multiplied by the factor $e^{cL(f)}$ where $c$ is the 
central charge and $L(f)$ is given by the Liouville formula

$$L(f)=\int _{\Sigma _{g}}[\frac{1}{2}df\wedge *df+fKG+\frac{1}{2}(e^{2f}-1)\omega _{G}]$$

where $G$ is the metric with Gauss curvature $K$ and volume form $\omega _{G}$.
Moreover $H_{S^{1}}$ carries a projective representation of the cobordism 
category, thus each cobordism $\Sigma _{g}$ has associated to it a complex line
 bundle (in its most general form these are tensor products of determinant line
 bundles which generalise the usual basis 
$L_{n}=e^{ih\theta}\frac{d}{d\theta}$ of $Vect(S^{1})_{\bf C}$).\\

This geometric picture of G. Segal is very nice but when trying to generalise
it in higher dimensions one faces problems.\\

Now we shall modify Segal's definition using the more convenient language
of \emph{operads}, we follow \cite{operads}. We shall swich to Hamiltonian 
formalism and point out that our formalism below is less satisfactory than 
G. Segal's because it works only for genus 0 Riemannian surfaces but it is 
easier to generalise in higher dimensions. We shal explain in detail
the structures appearing in the $D=2$ case and then we shall try to see how
much can be immediately generalised to higher dimensions.\\

Before doing that we would like to make a comment: there are \emph{two}
ways to construct operads related to Riemann surfaces: the first one is by
using moduli spaces of punctured Riemann surfaces and compactify them, such
operads are roughly denoted $M(n)$; the second is by decorating the puctures
with local coordinates. One can sew two Riemann surfaces together
unambiguously (up to modular equivalence) by using suitable local
coordinates.\\

Let $R(n)$ be the moduli space of nondegenerate Riemann spheres $\Sigma $
with $n$ labelled punctures and non-overlapping holomorphic discs at each
puncture (holomorphic embeddings of the standard disc $|z|< 1$ to $\Sigma $
centered at the puncture). The spaces $R(n)$, $n\in {\bf N}^{*}$ form an operad
under sewing Riemann spheres at punctures (cutting out the discs $|z|\leq r$
and $|w|\leq r$ for some $r=1-\epsilon $ at sewn punctures and identifying the
annuli $r < |z| < 1/r$ and $r < |w| < 1/r$ via $w=1/z$). The symmetric group
interchanges punctures along with the holomorphic discs.\\

Consider the complexification $V$ of the Virasoro algebra of complex valued
vector fields on the circle, generated by elements $L_{m}=z^{m+1}\partial /\partial z$, $m\in {\bf Z}$, with the commutators $[L_{m},L_{n}]=(n-m)L_{m+n}$.
Then one has:\\

{\bf Definition 7:}\\
A $D=2$ \emph{conformal field theory} at tree level consists of the following
data:
{\bf 1.} A topological vector space $H$ called \textsl{state space}.\\
{\bf 2.} An action $T:V\otimes H\rightarrow H$ of the Virasoro algebra $V$
on $H$.\\
{\bf 3.} A vector $|\Sigma >$ $\in Hom(H^{\otimes n}, H)$ for each $\Sigma \in
R(n)$ depending smoothly on $\Sigma $.
These data must satisfy the following relations:\\
{\bf 4.} $T(\underline{v})|\Sigma >=|\delta (\underline{v})\Sigma >$, where
$\underline{v}=(v_{1},...,v_{n})\in V$ and $\delta $ is the natural action of
$V^{n}$ on $R(n)$ by infinitesimal reparametrisations at punctures. In
particular $T(\underline{v})|\Sigma >=0$ whenever $\underline{v}$ can be
extended to a holomorphic vector field on $\Sigma $ outside of the discs.\\
{\bf 5.} The correspondence $\Sigma \mapsto |\Sigma >$ defines the structure
of an \emph{algebra over the operad} $R(n)$ on the space of states $H$.\\

So briefly, the slogan is that a $D=2$ CFT is an algebra over the operad
$R(n)$.\\

{\bf Definition 8:}\\
A \emph{string background} (at the tree level) is a $D=2$ CFT based on the
vector space $H$ with the following additional data:\\
{\bf 1.} A {\bf Z}-grading $H=\oplus _{i\in {\bf Z}}H_{i}$ on the state
space.\\
{\bf 2.} An action of the \textsl{Clifford algebra} $C(V\oplus V^{*})$ which
is denoted $b:V\otimes H\rightarrow H$ and $c:V^{*}\otimes H\rightarrow H$
for generators of the Clifford algebra, the degree of $b$ is -1 and the degree of $c$ is +1.\\
{\bf 3.} A differential $Q:H\rightarrow H$, $Q^{2}=0$, of degree +1 called
\emph{BRST operator} satisfying $Qb+bQ=T$.\\

The differential graded complex $(H,Q)$ is called the \emph{BRST complex} and
the degree is called the \textsl{ghost number}.\\

One of the nicest implications of a string background is the construction of
a morphism of complexes $\omega _{n}:Hom(H, H^{\otimes n})\rightarrow
\Omega ^{*}(R(n))$, from the complex of linear mappings between tensor powers
of the BRST complex $H$ to the de Rham complex of the space $R(n)$.\\

Taking the \textsl{cohomology} of the BRST complex
gives the \emph{space of physical states}. In physics this amounts to mod out
gauge invariance (this is a cohomological approach to \textsl{symplectic
reduction} in the case of a symplectic manifold  carrying a Lie group action
which is more convenient in infinite dimensions, i.e. field theory).\\

Then a \emph{closed string field theory} is a string background together with
a morphism of operads $\psi :M(n)\rightarrow R(n)$. This however does not
readily generalise to higher dimensions since both operads $M(n)$ and $R(n)$
are related to moduli spaces of Riemann surfaces. The aspect of closed string
field theory which will be useful for higher dimensional generalisations is
the following:\\

{\bf Key fact in physics:} The space of physical states in closed string field theory
(namely BRST cohomology) has the structure of a \emph{Gerstenhaber algebra}.\\

Let us explain this a little bit more: from the work of G. Segal we knew that
in general $D=2$ topological quantum
field theories (and in particular closed string field theory)
are ${\bf Z}$-graded commutative associative Frobenious algebras, hence they
are \textsl{graded
associative algebras}. Yet it was observed by Witten and Zwiebach that closed
string field theory also carries the structure of a \textsl{Differential
Graded Lie Algebra} relative to another grading which is
\emph{``the associative grading - 1''}.
In fact these 2 structures can be combined together to give a \emph{G-algebra}
structure.\\

Now we make use of Cohen's theorem saying that $G$-algebras
correspond exactly to the homology of the chains of the little 2-discs
operad and of Tamarkin's result to deduce that the \emph{space of physical
states of closed string field theory (which is a special case of $D=2$ CFT)}
has a natural (defined up to homotopy) \textsl{2-algebra structure}.\\

So this is the important relation between strings and the operad $C_{2}(n)$
that plays the fundamental role to understand Kontsevich's
statement which is then simply the \emph{higher dimensional version} of the
above fact which originally holds for strings ($D=2$ case).\\

\section{Discussion}

Let us start with some\\

{\bf Remarks:}\\

{\bf 1.} The original Deligne conjecture, namely the case $d=1$, was
proved by Kontsevich and Soibelman in 2000
using ideas and techniques from Dan Quillen's \emph{homotopical algebra} (see \cite{quillen}),
which roughly is a \textsl{non-linear}
generalisation of \emph{homological algebra}.\\

{\bf 2.} As Kontsevich explains in his article, from Deligne's conjecture
and from Tamarkin's theorem (namely the formality of the operad
$Chains C_{d}(n)$), follows almost immediately his
earlier result on deformation quantization of symplectic (Poisson) manifolds
for the case where the associative algebra of interest is just the polynomial
algebra in $n$ variables
$$A:={\bf R}[x_{1},...,x_{n}].$$
Let us recall that Kontsevich's result was that for the associative algebra
$A$ above one has that its Hochschild complex $C^{*}(A,A)$ is \emph{homotopic}
as a Lie algebra to its Hochschild cohomology $H^{*}(A,A)$. An equivalent
statement is that for the Euclidean space $X={\bf R}^{n}$, the Hochschild
complex of the associative algebra of functions on $X$ equipped with
the Gerstenhaber bracket is \emph{homotopic as DGLA} to
the ${\bf Z}$-graded superalgebra of \emph{polyvector fields} on $X$ equipped
with the Schouten-Nijenhuis bracket.\\

{\bf 3.} The structure of an $A_{\infty }$-algebra has appeared recently in
\emph{open strings}. Moreover let us mention that one of the main examples of
homotopy associative algebras (or $A_{\infty }$-algebras) is singular chains
of based loops.\\

Perhaps here we should mention another important
result: the BRST complex of closed string field theory has also the structure
of a $G_{\infty }$-algebra (see \cite{operads}). That makes someone to
speculate on a relation between the operads $G_{\infty }(n)$ and
$Chains C_{2}(n)$. For example it is unknown if double loop spaces (which are
the primary examples of 2-algebras) also carry a homotopy Gerstenhaber
algebra structure.\\

Now let us try to answer the following question: \emph{Why should physics care about 
the Deligne conjecture?} We think for 2 reasons:\\

{\bf i.} The fact that the Hochschild complex of an associative algebra is a
2-algebra (original conjecture) is related to the action
of the \emph{Grothendieck-Teichmuller (G-T) group}. The fact that the
Hochschild
complex plays the fundamental role in the theory of \emph{deformations} of
associative algebras explains why its study is important in quantum field
theory if one adopts the \emph{deformation quantization approach}.\\

In other
words the goal is to understand the \emph{action} of the G-T group on the
\emph{space of all
deformation quantizations} on the associative algebra of functions (in fact
one needs a little more structure, i.e. a Poisson algebra structure) on a
given (spacetime or phase space respectively) manifold and this is believed
to be related
to gauge symmetry. Yet all these are far from being clear at the moment.
Let us briefly recall that the Grothendieck-Teichmuller group can be defined
as the automorphism group of the tower of the pro-nilpotent completions of
the pure braid groups; the pure braid group of $n$ strings is the fundamental
group of the configuration space of $n$ points in the plane
${\bf F}(n, {\bf R^{2}})$.\\

{\bf ii.} Let us forget deformation quantization now and let's focus on 
holography: If one wants to understand $(d+1)$-dim \emph{quantum gravity}, one
approach is to follow holography. This means that equivalently one should 
understand
 $D=d$  CFT. From the string theory example we have learnt that $D=2$ CFT is
related to 2-algebras. It is then reasonable to expect that $D=d$ CFT should be
related to $d$-algebras. So the hope is that this higher dimensional Deligne
conjecture will tell us something about the \emph{``BRST complex''} whose 
cohomology would describe the $D=d$ CFT. So the $D=d$ CFT is a $d$-algebra 
and following the Hamiltonian formalism the space of physical states will be 
expresseed as a cohomology; then the Deligne conjecture will contain 
information about the \emph{cochain level} of the cohomology. This information at the cochain level has a nice 
\textsl{geometric interpretation} as follows (if one adopts the Lagrangian formalism of the theory):\\
 
Holography tells us that $(d+1)$-quantum gravity is reduced to some 
(conformal) field theory on the boundary. Obviously there are \emph{many} 
$(d+1)$-quantum gravity theories which will have the \emph{same} boundary 
theory, simply because there are \emph{many} $(d+1)$-manifolds with the \emph{same} 
$d$-manifold as boundary. Consider all $(d+1)$-bulk theories then with the 
same boundary $d$-theory as some sort of an \emph{extra ``gauge freedom''}, 
 ie that they define a ``fibre'' over the fixed boundary $d$-theory. Equivalently, we have a whole \emph{cobordismm class}
 of a boundary theory consisting of bulk theories which, because of holography, are \textsl{physically the same} theory.
Assuming 
that the boundary theory is described by a $d$-algebra, we know that the 
Grothendieck-Teichmuller group acts in an analogous way that the gauge group 
acts in ordinary gauge theories (the fibres are essentially the gauge orbits).
The higher dimensional Deligne conjecture is another way to express this action
 of the Grothendieck-Teichmuller group which takes care of this extra gauge 
freedom arising from the fact that many $(d+1)$-bulk theories have the same 
resulting boundary $d$-theory. This is roughly the statement that the 
Hochschild complex of a $d$-algerba has a natural (appropriately defined up to 
homotopy) $(d+1)$-algebra structure.\\ 

Yet this is not still the end of the story: the action of the G-T group is 
closely related to the action of the \emph{motivic group}: since the little
$d$-discs operad is formal, we can form a \emph{torsor}; in general given any 
pair of equivalent objects $A$ and $B$ in a category, we can form the space
of all isomorphisms between the objects $A$ and $B$ denoted $Iso(A,B)$. The 
groups of automorphisms of both $A$ and $B$ denoted $Aut(A)$ and $Aut(B)$ 
respectively act naturally on $Iso(A,B)$ and their actions commute. A torsor
is a structure that encodes this information. If we think of the formality of
the operad $C_{d}(n)$ as defining a pair of equivalent objects in the category
of ${\bf Z}$-graded complexes, we can indeed define a torsor and then the 
action of the G-T group will play the role of say $Aut(A)$ whereas the motivic 
group wil play the role of $Aut(B)$. Further evidence that our proposal to use
$d$-algebras to describe $D=d$ CFT is not at all unreasonable comes from the 
fact that Tamarkin in order to prove his formality theorem he made use of the
Drinfeld-Kazhdan associator whose origin is the Knizhnik-Zamolodchikov 
equations, namely it comes directly from $D=2$ CFT! (For the precise 
definitions of the motivic group and the Drinfeld-Kazhdan associator we refer 
to \cite{kon1}).\\

Going back to physics a pessimist might argue that the holography principle 
does not really
``resolve'' the BH paradoxe; it simply pushes it even further to the realm
 of quantum gravity which is \emph{terra incognita} for today's physics.
We however would like to adopt a more positive point of view: by accepting
the validity of holography we can actually
\emph{use} it in order to learn something about quantum gravity and
at the same time we try to understand its wider implications for physics.
Finally we would like to mention that there have been already some positive
tests for the validity of holography mainly in the framework of some
calculations related to its string theoretic version (Maldacena conjecture).\\

Operads (and the little discs operad in particular) currently should be seen 
more as an algebraic framework to conceptually understand higher dimensional 
CFT's. As a computational device in $D=2$ CFT people use Topologiacl Vertex 
Operator Algebras (TVOA) as a tool to construct topological CFT but again their
 higher dimensional generalisation is not obvious.\\

The final comment we would like to make is the following: the structure of a 
Gerstenhaber algebra which was observed in $D=2$ CFT ment that the dot product
(which gives the OPE) and the Lie bracket (which gives the conformal symmetry)
are combined in a ``nice way'' in the space of physical states (ie that the 
bracket is a derivation with respect to the product). Our proposal that higher 
dimensional CFT's are described by $d$-algebras means that roughly we 
still have this nice combination in higher dimensions between the dot product 
and the Lie bracket. However in  general a Gerstenhaber algebra 
``does not care'' whether the Lie bracket comes specifically from conformal 
symmetry. Thus our proposal looks more like an algebraic framework to 
understand higher dimensional CFT and holography for the moment; however 
if we want to make its relation with physics---with conformal symmetry in 
particular---more 
concrete and if we want to properly and fully justify the proposal that a 
$d$-algebra should describe $D=d$ CFT, we would have to add more specific 
information about conformal 
symmetry (Lie bracket) and dot product (OPE) in higher dimensions. So it seems 
that the problem is not whether our proposal is valid but rather that it may be
  far too 
general to be of any practical use. In any case it helps conceptually we 
believe. For example one desired property would be to consider representations 
of the operad $C_{d}(n)$ on the endomorphism operad of a \textsl{Clifford 
algebra} (and not just a vector space) in order to make the theory 
\textsl{chiral}. However for the 
moment we do not know how to incorporate the \emph{Cardy condition} in our 
formalism.\\
 
{\bf Acknowledgements:}\\

The author would like to thank \textsl{Maxim Kontsevich} and \textsl{Graeme Segal} for useful 
discussions. This work was partly 
supported financially by an E.P.S.R.C. research grant (contract No: GR/R64995/01). An earlier version of this article under 
the title 
\textsl{"Holography and the Deligne Conjecture"} was presented as a poster/oral presentaton during the 
\emph{XIVth International Congress of Mathematical Physics}, Lisbon, Portugal, 28 July-2 Agust 2003.\\

\end{document}